\documentclass[11pt,draftclsnofoot,onecolumn]{IEEEtran}
\usepackage{setspace}
\usepackage{geometry}
\usepackage{graphicx}
\usepackage{amsmath}
\usepackage{multirow}
\usepackage{fancyhdr}

\def\II{{{\mathbf{I}}}}

\begin{document}
\doublespace

\fancyhf{} % clear all header and footer fields
\fancyhead[L]{\scriptsize\textit{This paper is a preprint of a paper accepted by IET Information Security and is subject to Institution of Engineering and Technology Copyright. When the final version is published, the copy of record will be available at IET Digital Library.}}

\title{Security and complexity of the McEliece cryptosystem based on QC-LDPC codes
\thanks{This work was supported in part by the MIUR project ``ESCAPADE''
(Grant RBFR105NLC) under the ``FIRB -– Futuro in Ricerca 2010'' funding program.}
}

\author{Marco Baldi, Marco Bianchi and Franco Chiaraluce\\
Dipartimento di Ingegneria dell'Informazione,\\
Universit\`a Politecnica delle Marche,\\
Ancona, Italy;\\
E-mail: \{m.baldi, m.bianchi, f.chiaraluce\}@univpm.it}

\maketitle

\thispagestyle{fancy}

\begin{abstract}
In the context of public key cryptography, the McEliece cryptosystem 
represents a very smart solution based on the hardness of the decoding 
problem, which is believed to be able to resist the advent of
quantum computers.
Despite this, the original McEliece cryptosystem, based on Goppa
codes, has encountered limited interest in practical applications, partly 
because of some constraints imposed by this very special class of codes.
We have recently introduced a variant of the McEliece cryptosystem
including low-density parity-check codes, that are state-of-the-art
codes, now used in many telecommunication standards and applications.
In this paper, we discuss the possible use of a bit-flipping 
decoder in this context, which gives a significant advantage in
terms of complexity.
We also provide theoretical arguments and practical tools for 
estimating the trade-off between security and complexity, 
in such a way to give a simple procedure for the system design.
\end{abstract}

\section{Introduction}

In recent years, a renewed interest has been devoted to the McEliece cryptosystem \cite{McEliece1978},
which is one of the most attractive options for \textit{post-quantum} public key cryptography. It exploits
error correcting codes to obtain both the private and the public key.
Its security relies on the difficulty of decoding a linear block code without any known structure. More precisely, two kinds of attacks can be mounted against this system. The former aims at retrieving the private key from the public key, while the latter tries to recover the cleartext from the ciphertext, without the knowledge of the private key.
The first kind of attack can be avoided through a suitable choice of the codes to be used in the system, and of their parameters.
The second kind of attack basically consists in decrypting the intercepted ciphertext without knowing the private key.
This can be achieved by using information set decoding algorithms on the public code.
Algorithms of this kind have been investigated since a long time \cite{McEliece1978, Lee1988}.
Other approaches exploit improvements of the probabilistic algorithm first proposed by Stern in \cite{Stern1989}.
These are presented in \cite{Canteaut1998} and, more recently, in \cite{Bernstein2008, Peters2010, May2011, Bernstein2011, Becker2012}.

The original version of the McEliece cryptosystem, based on binary Goppa codes 
with irreducible generator polynomials, is faster than the widespread RSA cryptosystem. 
However, it has two major drawbacks: large keys and low transmission rate, the latter 
being coincident with the code rate.
The McEliece cryptosystem uses generator matrices and encodes the messages into codewords
of the public code.
Niederreiter proposed a code-based cryptosystem using the parity-check matrix and Generalized Reed--Solomon
(GRS) codes \cite{Niederreiter1986}. This proposal was broken by Sidelnikov and Shestakov \cite{Sidelnikov1992};
however, it still works with Goppa codes, as shown in \cite{Li1994}.
The main advantage of Niederreiter's variant, which encodes the messages into syndrome vectors,
is to achieve a significant reduction in the number of operations for encryption, though this
is paid with a moderate increase in the number of operations for decryption.

The most effective way to overcome the drawbacks of the McEliece cryptosystem
is to replace Goppa codes with other families of codes, yielding a more compact 
representation of their characteristic matrices, 
and permitting to increase the code rate.
Unfortunately, although several families of codes with such characteristics exist,
it is very difficult to replace Goppa codes with other codes without incurring into
serious security flaws, as occurred, for example, with Gabidulin codes \cite{Overbeck2008}
and GRS subcodes \cite{Wieschebrink2010}.

Among the most recent proposals, Quasi-Cyclic (QC) \cite{Berger2009}, Quasi-Dyadic (QD) 
\cite{Misoczki2009} and Quasi-Cyclic Low-Density Parity-Check (QC-LDPC) codes \cite{Baldi2009}
have been considered for possible inclusion in the McEliece cryptosystem and also in 
symmetric key secure channel coding schemes \cite{SobhiAfshar2009}.
However, the solutions \cite{Berger2009} and \cite{Misoczki2009} have been successfully attacked in \cite{Umana2010} and \cite{Faugere2010}.
An updated variant of the quasi-dyadic solution has been recently proposed in \cite{Barbier2011}, and it should be more secure; however, the complexity of the attack in \cite{Faugere2010}
for the binary QD case is still open, and work is in progress on such issue.
The attack procedure described in \cite{Faugere2010, Faugere2010a} exploits
an algebraic approach, based on a system of bi-homogeneous polynomial equations, which
holds for the whole class of alternant codes. Hence, such attack concerns all
cryptosystems using codes in this family.

LDPC codes are state-of-art error correcting codes, first introduced by Gallager
in the sixties \cite{Gallager}, and more recently rediscovered \cite{Tanner1981, MacKay99, Richardson2003RenaissanceLDPC}.
While random-based LDPC codes are able to approach the channel capacity \cite{Richardson2001},
structured LDPC codes have the advantage of an easier implementation
of the encoding and decoding operations, and benefit from reduced storage requirements \cite{Tanner2001}.
QC-LDPC codes are one of the most important examples of structured LDPC codes, and they
have also been proved to achieve very good performance \cite{Chen2004}.
The existence of efficient iterative decoding algorithms for LDPC codes is the distinguishing
feature of this class of codes. The rationale of these algorithms is an iterated updating and
exchange of messages along a bipartite graph, also known as Tanner graph, which represents the
code parity-check matrix.
Very good decoding performance is achieved as long as the code Tanner graph is free of short
cycles, that is, closed loops starting and ending at one node.
More details on LDPC decoding will be given in Section \ref{sec:EncDec}.

Concerning the use of LDPC codes in the McEliece cryptosystem, they were initially thought 
to be unable to give significant advantages, due to the fact that the sparse nature of their matrices cannot be 
exploited for reducing the key size \cite{Monico2000}. Furthermore, adopting very
large codes was found to be necessary for avoiding that the intrinsic code sparsity is 
exploited by an attack to the dual of the public code \cite{Baldi2007ICC}.
However, it has also been shown that, by replacing the permutation matrix used for
obtaining the public key with a more general transformation matrix, the code sparsity
can be hidden and the attack to the dual code avoided \cite{Baldi2007ISIT}.
Unfortunately, the proposal in \cite{Baldi2007ISIT} still used only sparse transformations, 
which exposed it to a total break attack \cite{Otmani2008}.
Subsequently, however, we have presented a simple modification that allows 
to avoid such flaw, so obtaining a QC-LDPC 
code-based cryptosystem that is immune to any known attack \cite{Baldi2008}.
This version of the cryptosystem is able to reduce the key size with respect 
to the original version and also to use higher code rates, which is in line
with the most recent proposals concerning McEliece variants.
Moreover, the size of its public keys increases linearly with the code dimension;
so, it scales favorably when larger keys are needed for facing the increasing computing
power.

In this paper, we elaborate on our last proposal, first by describing bit-flipping
decoding \cite{Gallager1963} for the considered QC-LDPC codes,
which yields a 
significant reduction in the decoding complexity, at the cost of a moderate loss in 
terms of error correction.
The performance of bit-flipping decoding can be easily predicted 
through theoretical arguments, and this helps dimensioning the system, without 
the need of long numerical simulations.
We also consider the most effective attack procedures known up to now and 
estimate analytically their work factor (WF).
This way, we provide tools that allow to easily
find the best set of system parameters aiming at optimizing
the trade-off between security and complexity.

The paper is organized as follows: in Section \ref{sec:System}, we describe
the proposed version of QC-LDPC code-based cryptosystem; in Section \ref{sec:EncDec},
we describe the encryption and decryption algorithms and evaluate their complexity;
in Section \ref{sec:Security}, we assess the security level of the system;
finally, Section \ref{sec:Conclusion} concludes the paper.

\section{McEliece cryptosystem based on QC-LDPC codes}
\label{sec:System}
\begin{figure}
\begin{centering}
\includegraphics[keepaspectratio, width=120mm]{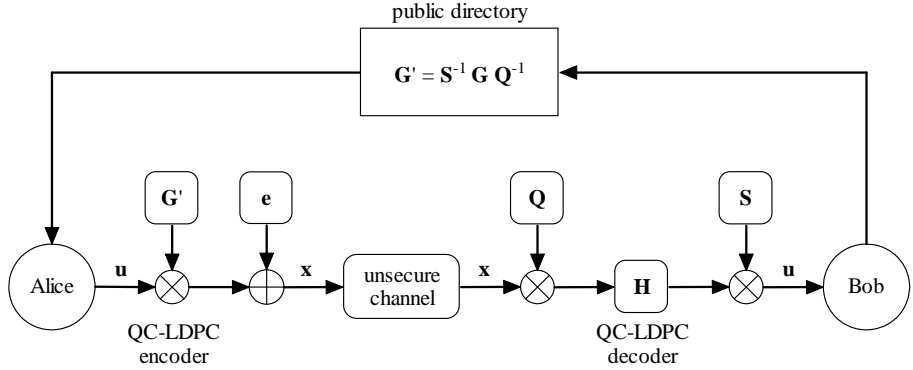}
\caption{The McEliece cryptosystem based on QC-LDPC codes.}\label{fig:System}
\par\end{centering}
\end{figure}
The main functions of the McEliece cryptosystem based on QC-LDPC codes are
shown in Fig. \ref{fig:System}: QC-LDPC codes with length $n = n_0 \cdot p$, dimension 
$k = k_0 \cdot p$ and redundancy $r = p$ are adopted, where $n_0$ is a small integer
(e.g., $n_0 = 3,4$), $k_0 = n_0-1$, and $p$ is a large integer (on the order of some
thousands).
For fixed values of the parameters, the private key is formed by the sparse 
parity-check matrix $\mathbf{H}$ of one of these codes, randomly chosen, having 
the following form:
\begin{equation}
\mathbf{H} = \left[ \mathbf{H}_{0} | \mathbf{H}_{1} | \ldots |\mathbf{H}_{n_0-1} \right],
\label{eq:HCircRow}
\end{equation}
that is, a row of $n_0$ circulant blocks $\mathbf{H}_{i}$, each with row 
(column) weight $d_v$.
Without loss of generality, we can suppose that $\mathbf{H}_{n_0-1}$ is non 
singular; so, a systematic generator matrix for the code is $\mathbf{G} = [\mathbf{I} | \mathbf{P}]$, where $\mathbf{I}$ represents the $k \times k$ identity matrix and
\begin{equation}
\mathbf{P}=\left[\begin{array}{c}
\left(\mathbf{H}_{n_{0}-1}^{-1}\cdot\mathbf{H}_{0}\right)^{T}\\
\left(\mathbf{H}_{n_{0}-1}^{-1}\cdot\mathbf{H}_{1}\right)^{T}\\
\vdots\\
\left(\mathbf{H}_{n_{0}-1}^{-1}\cdot\mathbf{H}_{n_{0}-2}\right)^{T}\end{array}\right]
\label{eq:GCirc}
\end{equation}
where superscript $^T$ denotes transposition.
Concerning the computation of $\mathbf{H}_{n_{0}-1}^{-1}$, we observe that the inverse
of a circulant matrix can be computed through techniques which are significantly more 
efficient than naive inversion \cite{Baldi2011a}.

Let us denote by $\mathbf{h}_i$, $i = 0 \ldots n_0-1$, the vector containing the 
positions of symbols $1$ in the first row of the matrix $\mathbf{H}_i$, $i = 0 \ldots n_0-1$.
It is easy to show that, if all the $\mathbf{h}_i$ vectors 
have disjoint sets of differences modulo $p$, the matrix $\mathbf{H}$ is free of length-$4$ 
cycles in its associated Tanner graph.
%, which is an essential prerequisite for effective LDPC decoding.
The secret code can be easily constructed by randomly selecting
$n_0$ vectors $\mathbf{h}_i$ with such property.
This permits us to obtain large families of codes with identical parameters \cite{Baldi2007ISIT}.
Under the LDPC decoding viewpoint, most of the codes in a family have the same 
properties; so, they show comparable error correction performance when \textit{belief propagation} 
\cite{Pearl1982} decoding algorithms are adopted.

In the QC-LDPC code-based cryptosystem, Bob chooses a secret QC-LDPC code
by generating its parity-check matrix, $\mathbf{H}$, and chooses 
two more secret matrices: a $k \times k$ non singular scrambling matrix $\mathbf{S}$ 
and an $n \times n$ non singular transformation matrix $\mathbf{Q}$ with 
row/column weight $m$.
Then, he obtains a systematic generator matrix $\mathbf{G}$ for the secret 
code, in the form $\mathbf{G} = [\mathbf{I} | \mathbf{P}]$, and produces his public key as:
\begin{equation}
\mathbf{G}' = \mathbf{S}^{-1} \cdot \mathbf{G} \cdot{\mathbf{Q}^{-1}}.
\end{equation}
%We notice that 
The public key is a dense matrix, but,
%; so, apparently, the sparse 
%character of LDPC codes does not help reducing the key length. 
%However, 
since we adopt QC-LDPC codes,
%the characteristic matrices are formed by circulant blocks, that are completely described by one of their rows. Thus,
the knowledge of one row of each circulant block is sufficient to describe it. % specify the whole matrix.
We notice that, differently from the original McEliece cryptosystem,
the public code is not permutation-equivalent to the private code.
In fact, the permutation matrix used in the original system \cite{McEliece1978} has been
replaced by $\mathbf{Q}$, that is a sparse $n \times n$ matrix, with row 
and column weight $m>1$.
This way, the LDPC matrix of the secret code ($\mathbf{H}$) is mapped into 
a new parity-check matrix valid for the public code:
\begin{equation}
\mathbf{H}' = \mathbf{H} \cdot \mathbf{Q}^T
\end{equation}
and, through a suitable choice of $m$, the density of $\mathbf{H}'$ can be 
made high enough to avoid attacks to the dual code.

Alice %, in order to send encrypted messages to Bob, 
fetches 
$\mathbf{G}'$ from the public directory, divides her message into $k$-bit
words, and applies the encryption map as follows:
\begin{equation}
\mathbf{x} = \mathbf{u} \cdot \mathbf{G}' + \mathbf{e},
\end{equation}
where $\mathbf{x}$ is the ciphertext corresponding to the cleartext $\mathbf{u}$,
and $\mathbf{e}$ is a random vector of $t'$ intentional errors.
After receiving $\mathbf{x}$, Bob inverts the transformation as follows:
\begin{equation}
\mathbf{x}' = \mathbf{x} \cdot \mathbf{Q} = \mathbf{u} \cdot \mathbf{S}^{-1} \cdot \mathbf{G} + \mathbf{e} \cdot \mathbf{Q},
\label{Eq:six}
\end{equation}
thus obtaining a codeword of the secret LDPC code affected by
the error vector $\mathbf{e} \cdot \mathbf{Q}$, with weight $\leq t=t'm$.
%Then,
Bob must be able to correct all the errors through LDPC decoding 
and to obtain $\mathbf{u}' = \mathbf{u} \cdot \mathbf{S}^{-1}$.
Finally, he can recover $\mathbf{u}$ from $\mathbf{u}'$, through multiplication by $\mathbf{S}$.

We note from (\ref{Eq:six}) that the introduction of the matrix $\mathbf{Q}$ 
%in place of the permutation matrix
causes an error propagation effect 
(at most by a factor $m$) within each received frame. 
This is compensated by the high error correction capability of the
%secret
QC-LDPC code, that must be able to correct up to $t$ errors.
Suitable QC-LDPC codes can be designed for such purpose. However, we must also note that, contrary to the McEliece cryptosystem based on Goppa codes, which corrects all errors of a certain prescribed weight, the decoding radius of LDPC codes is usually unknown. So, there is a small probability that Bob fails to recover the secret message.
To prevent such event, different procedures can be implemented. First, Bob can make a careful selection of the private code, rather than just picking up the first code randomly generated. In fact, the number of codes that can be obtained through random-based approaches, like random difference families \cite{Baldi2007ISIT}, is impressively high.
Secondly, when the cryptosystem is used for data transmissions, an automatic repeat request (ARQ) protocol can allow Alice to know whether Bob is able to correct all the errors she has randomly introduced or not.
Indeed, Bob is able to detect uncorrected frames through the parity check performed by the LDPC decoder, and, consequently, he can request retransmission. In this case, a new random vector is generated by Alice, and the procedure is repeated until a correctable error pattern is obtained.
In principle, this exposes the system to message-resend attacks, but simple modifications of the cryptosystem are known which prevent these attacks without significant drawbacks \cite{Sun1998, Engelbert2007}.
As will be observed in the next section, using these conversions is also advantageous from the key size standpoint.
Obviously, this additional effort increases the latency, but the problem is not serious if
the number of errors is properly chosen and controlled.

\section{Encryption, decryption and their complexity}
\label{sec:EncDec}

%In this section, we describe the operations required to perform encryption and decryption in the QC-LDPC code-based cryptosystem.

%After having identified what algorithms are best suited to effectively implement the required operations, we provide arguments for estimating their complexity as a function of the system parameters.

\subsection{Key size and transmission rate}
\label{subsec:KeySize}

%A first characteristic to be investigated for the QC-LDPC code-based cryptosystem is the size of its public keys and the transmission rate, that are the two main aspects to be improved with respect to the original McEliece cryptosystem.

In the QC-LDPC code-based cryptosystem, due to the special form \eqref{eq:HCircRow} of the matrix $\mathbf{H}$, the
code rate is $(n_0-1)/n_0$.
In the following, we will focus on two values of $n_0$, namely: $n_0 = 3,4$, which 
give transmission rates equal to $2/3$ and $3/4$, respectively.

%As for the key size, it coincides with the amount
%of data needed to know $\mathbf{G}'$, that is, 

Concerning the key size, we observe that, in the considered system, 
the public key is a binary matrix formed by $k_0 \times n_0 = (n_0 - 1) \times n_0$ 
circulant blocks, each with size $p \times p$. 
Since each circulant block is completely described by a single row (or column), that is, 
$p$ bits, the public key size is $k_0 \cdot n_0 \cdot p = (n_0 - 1) \cdot n_0 \cdot p$ bits.

This size can be further reduced if we consider that a suitable conversion
is needed to make the McEliece cryptosystem secure against some classical
attacks, like partial message knowledge and message resend attacks \cite{Engelbert2007}.
Attacks of this kind can be avoided by using some CCA2-secure variants of the McEliece
cryptosystem, which have in common the idea of scrambling the input messages.
When these variants are used, the public matrix can be put in systematic
form \cite{Bernstein2008}, so the memory needed to store it becomes $k_0 \cdot (n_0 - k_0) \cdot p = (n_0 - 1) \cdot p$ bits.

The values of the key size (expressed in bytes), estimated considering the use of a CCA2-secure variant, are reported in Table \ref{tab:KeySize}, for $n_0 = 3,4$
and for a set of values of $p$ we will consider throughout the paper.
%As we observe from the table,
%ignoring for the moment the security issue (that will be assessed in the following),
All choices of the system parameters we have considered give smaller key size and higher
transmission rate than those of the original McEliece cryptosystem \cite{McEliece1978} and 
its Niederreiter version \cite{Niederreiter1986}.
Considering a CCA2-secure conversion, both they have a key length of $32750$ bytes, and 
rate $0.51$ and $0.57$, respectively.

\begin{table}[ht]
\renewcommand{\arraystretch}{1.3}
\caption{Public key size expressed in bytes.}
\label{tab:KeySize}
\centering
\small\begin{tabular}{|c||@{\hspace{1mm}}c@{\hspace{1mm}}|@{\hspace{1mm}}c@{\hspace{1mm}}|@{\hspace{1mm}}c@{\hspace{1mm}}|@{\hspace{1mm}}c@{\hspace{1mm}}|@{\hspace{1mm}}c@{\hspace{1mm}}|@{\hspace{1mm}}c@{\hspace{1mm}}|@{\hspace{1mm}}c@{\hspace{1mm}}|@{\hspace{1mm}}c@{\hspace{1mm}}|@{\hspace{1mm}}c@{\hspace{1mm}}|@{\hspace{1mm}}c@{\hspace{1mm}}|@{\hspace{1mm}}c@{\hspace{1mm}}|@{\hspace{1mm}}c@{\hspace{1mm}}|@{\hspace{1mm}}c@{\hspace{1mm}}|}
\hline
%& \multicolumn{13}{c|}{$p$ [bits]} \\
%\hline
$p$ [bits] & $4096$ & $5120$ & $6144$ & $7168$ & $8192$ & $9216$ & $10240$ & $11264$ & $12288$ & $13312$ & $14336$ & $15360$ & $16384$ \\
\hline
\hline
%$n_0=3$ & $3072$ & $3840$ & $4608$ & $5376$ & $6144$ & $6912$ & $7680$ & $8448$ & $9216$ & $9984$ & $10752$ & $11520$ & $12288$ \\
$n_0=3$ & $1024$ & $1280$ & $1536$ & $1792$ & $2048$ & $2304$ & $2560$ & $2816$ & $3072$ & $3328$ & $3584$ & $3840$ & $4096$ \\
\hline
%$n_0=4$ & $6144$ & $7680$ & $9216$ & $10752$ & $12288$ & $13824$ & $15360$ & $16896$ & $18432$ & $19968$ & $21504$ & $23040$ & $24576$ \\
$n_0=4$ & $1536$ & $1920$ & $2304$ & $2688$ & $3072$ & $3456$ & $3840$ & $4224$ & $4608$ & $4992$ & $5376$ & $5760$ & $6144$ \\
\hline
\end{tabular}
\end{table}

\subsection{Multiplication by circulant matrices}
\label{subsec:Winograd}

%Both the private and the public key of the considered cryptosystem are
%given by two matrices formed by circulant blocks.
A fundamental point for reducing complexity in the considered cryptosystem
is to adopt efficient algorithms for performing multiplication of a
circulant matrix by a vector.

Since circulant matrices are also Toeplitz matrices, an effective algorithm
for fast computation of vector-matrix products is the Winograd convolution 
\cite{Winograd}.
The Winograd algorithm is a generalization of the Karatsuba-Ofman algorithm,
that has been reviewed even recently, in the perspective to allow fast VLSI 
implementations \cite{Fan2010}.
If we consider a $p \times p$ Toeplitz matrix $\mathbf{T}$, with even $p$,  
we can decompose it as follows:
\begin{equation}
\mathbf{T} = \begin{bmatrix}
\mathbf{T}_0&\mathbf{T}_1\\\mathbf{T}_2&\mathbf{T}_0
\end{bmatrix}
=
\begin{bmatrix}
\II&\mathbf{0}&\II\\
\mathbf{0}&\II&\II
\end{bmatrix}
\begin{bmatrix}
\mathbf{T}_1-\mathbf{T}_0&\mathbf{0}&\mathbf{0}\\
\mathbf{0}&\mathbf{T}_2-\mathbf{T}_0&\mathbf{0}\\
\mathbf{0}&\mathbf{0}&\mathbf{T}_0
\end{bmatrix}
\begin{bmatrix}
\mathbf{0}&\II\\
\II&\mathbf{0}\\
\II&\II
\end{bmatrix}\ ,
\label{eq:WinogradMult}
\end{equation}
where $\II$ and $\mathbf{0}$ are the $p/2 \times p/2$ identity and null matrices, respectively, 
and $\mathbf{T}_0, \mathbf{T}_1, \mathbf{T}_2$ are $p/2 \times p/2$ Toeplitz matrices, as well 
as $\mathbf{T}_1-\mathbf{T}_0$ and $\mathbf{T}_2-\mathbf{T}_0$.
It follows that the multiplication of a vector $\mathbf{V} = \left[\mathbf{V}_0 \ \mathbf{V}_1\right]$
by the matrix $\mathbf{T}$ can be split into three phases:

\begin{itemize}
\item \textit{Evaluation phase}: multiplication of $\mathbf{V}$ by the first matrix translates into
the addition of two $p/2$-bit vectors ($\mathbf{V}_0$ and $\mathbf{V}_1$); so, its cost, in terms of
binary operations, is $p/2$.
\item \textit{Multiplication phase}: the vector resulting from the evaluation phase must be multiplied
by the second matrix. This translates into $3$ vector-matrix products by $p/2 \times p/2$ Toeplitz matrices.
If $p/2$ is even, the three multiplications can be computed in a recursive way, by splitting each of
them into four $p/4 \times p/4$ blocks. If $p/2$ is odd (or sufficiently small to make splitting no
more advantageous), the vector-matrix multiplication can be performed in the traditional way and its
complexity is about $\left(p/2\right)^2/2$.
\item \textit{Interpolation phase}: the result of the multiplication phase must be multiplied by
the third matrix. This requires $2$ additions of $p/2$-bit vectors, that is, further $p$ binary 
operations.
\end{itemize}

The matrix $\mathbf{G}'$ used in the QC-LDPC code-based cryptosystem is formed 
by $k_0 \times n_0$ circulant blocks with size $p \times p$. 
When a vector is multiplied by such matrix, we can split the vector into $k_0$-bit subvectors and 
consider $k_0 \cdot n_0$ vector-matrix multiplications.
However, we must take into account that the evaluation phase on the $k_0$-bit subvectors must
be performed only once, and that further $(k_0 - 1) \cdot n_0 \cdot p$ binary operations are
needed for re-combining the result of multiplication by each column of circulants.

\subsection{Encryption operations and complexity}
\label{subsec:Encryption}

Encryption is performed by calculating the product 
$\mathbf{u}\cdot\mathbf{G}'$ and then adding the intentional error vector $\mathbf{e}$.
So, the encryption complexity can be estimated by considering the cost of a vector-matrix
multiplication through the Winograd convolution and adding $n$ binary operations for
summing the intentional error vector.

Table \ref{tab:Encryption} reports the values of the encryption complexity, expressed
in terms of the number of binary operations needed for each encrypted bit, as a function
of the circulant matrix size $p$, for $n_0 = 3$ and $n_0 = 4$.
The use of the Winograd convolution is particularly efficient when $p$ is a power
of $2$, since, in such cases, recursion can be exploited to the utmost.

\begin{table}[ht]
\renewcommand{\arraystretch}{1.3}
\caption{Binary operations needed for each encrypted bit.}
\label{tab:Encryption}
\centering
\small\begin{tabular}{|c||@{\hspace{1mm}}c@{\hspace{1mm}}|@{\hspace{1mm}}c@{\hspace{1mm}}|@{\hspace{1mm}}c@{\hspace{1mm}}|@{\hspace{1mm}}c@{\hspace{1mm}}|@{\hspace{1mm}}c@{\hspace{1mm}}|@{\hspace{1mm}}c@{\hspace{1mm}}|@{\hspace{1mm}}c@{\hspace{1mm}}|@{\hspace{1mm}}c@{\hspace{1mm}}|@{\hspace{1mm}}c@{\hspace{1mm}}|@{\hspace{1mm}}c@{\hspace{1mm}}|@{\hspace{1mm}}c@{\hspace{1mm}}|@{\hspace{1mm}}c@{\hspace{1mm}}|@{\hspace{1mm}}c@{\hspace{1mm}}|}
\hline
%& \multicolumn{13}{c|}{$p$ [bits]} \\
%\hline
$p$ [bits] & $4096$ & $5120$ & $6144$ & $7168$ & $8192$ & $9216$ & $10240$ & $11264$ & $12288$ & $13312$ & $14336$ & $15360$ & $16384$ \\
\hline
\hline
$n_0=3$ & $726$ & $823$ & $919$ & $1005$ & $1092$ & $1178$ & $1236$ & $1351$ & $1380$ & $1524$ & $1510$ & $1697$ & $1639$ \\
\hline
$n_0=4$ & $956$ & $1081$ & $1206$ & $1321$ & $1437$ & $1552$ & $1624$ & $1783$ & $1811$ & $2013$ & $1984$ & $2244$ & $2157$ \\
\hline
\end{tabular}
\end{table}

The values reported in Table \ref{tab:Encryption} refer to the case of a non-systematic $\mathbf{G}'$,
that is, a generator matrix formed by $n_0 - 1 \times n_0$ generic circulant matrices.
Actually, when a CCA2-secure variant of the system is used, $\mathbf{G}'$ can be put in
systematic form, and, in this case, only $n_0 - 1$ vector-matrix multiplications are needed,
for the non-systematic part.
However, to implement a CCA2-secure variant, some suitable scrambling operation must be 
performed on the message, before multiplication by $\mathbf{G}'$.
In this case, the complexity depends on the chosen variant, and becomes more involved to estimate.
Since message scrambling followed by systematic coding is approximately equivalent
to non-systematic coding, we prefer to consider the latter, which allows for a straightforward
complexity estimation.

\subsection{Decryption operations and complexity}
\label{subsec:Decryption}

%In the considered cryptosystem,
Bob must perform the following three operations for decrypting the received message:
\begin{enumerate}
\item calculate the product $\mathbf{x}\cdot\mathbf{Q}$;
\item decode the secret LDPC code;
\item calculate the product $\mathbf{u}'\cdot\mathbf{S}$.
\end{enumerate}
Matrices $\mathbf{Q}$ and $\mathbf{S}$ are formed, respectively, by $n_0 \times n_0$ 
and $k_0 \times k_0$ circulant blocks. However, while the matrix $\mathbf{S}$ is dense,
the matrix $\mathbf{Q}$ is sparse (with row/column weight $m \ll n$).
So, it is advantageous to use the traditional multiplication (requiring $n \cdot m$ binary operations)
for calculating the product $\mathbf{x}\cdot\mathbf{Q}$.
On the contrary, the complexity of step 3) can be reduced by resorting to the
Winograd convolution for efficient multiplication of a vector by a circulant matrix.
Concerning step 2), Bob must exploit the secret LDPC matrix to implement a suitable 
decoding algorithm for trying to correct all intentional errors (that are $\leq t = t'm$).
LDPC decoding is usually accomplished through iterative decoding algorithms, which work on the code Tanner graph,
and implement the belief propagation principle to provide very good error correction capability.
Among them: the sum-product algorithm (SPA) \cite{Hagenauer1996} and the bit-flipping
(BF) algorithm \cite{Gallager1963}.
The SPA exploits real valued messages and ensures the best performance on channels with 
soft information, though with some dependence on finite precision issues \cite{Baldi2009a}.
When soft information from the channel is not available, as it occurs in our case,
it may be advantageous to use
the BF algorithm, which works on binary messages and requires very low complexity,
though its performance is not as good as that of the SPA.

The principle of the BF algorithm was devised in Gallager's seminal work
for LDPC codes with a tree representation \cite{Gallager1963}.
Given an LDPC parity-check matrix with column weight $d_v$, the 
variable nodes of its Tanner graph are initially filled in with the received codeword bits.
During an iteration, every check node $c_i$ sends each neighboring variable node $v_j$
the binary sum of all its neighboring variable nodes other than $v_j$.
So, each variable node receives $d_v$ parity-check sums. In order to send a message back
to each neighboring check node $c_i$, the node $v_j$ counts the number of unsatisfied
parity-check sums from check nodes other than $c_i$.
Let us denote by $b \leq d_v-1$ a suitably chosen integer; if the number of unsatisfied parity-check sums counted by $v_j$ is greater than or equal to $b$, then $v_j$ flips its value and sends it to $c_i$; otherwise, it sends its initial value 
unchanged to $c_i$. At the next iteration, the check sums are updated with such new values, until all of
them are satisfied or a maximum number of iterations is reached. 

A relevant issue concerns the choice of $b$. Two algorithms, commonly named A and B, were originally proposed by Gallager \cite{Gallager1963}: in Algorithm A, the value is fixed to $b = d_v-1$, while in Algorithm B it can vary between $\left\lceil d_v/2 \right\rceil$
and $d_v-1$ during decoding ($\left\lceil \cdot \right\rceil$ is the ceiling function). 
Algorithm A is simpler to implement, but Algorithm B ensures better performance.

%If such number is $\geq b \leq d_v-1$,
%then $v_j$ flips its value and sends it to $c_i$, otherwise $v_j$ sends its initial value 
%unchanged to $c_i$.

%At the next iteration, the check sums are updated with such new values, until all of
%them are satisfied or a maximum number of iterations is reached. 
%Two algorithms, named A and B, were originally proposed by Gallager \cite{Gallager1963}: in algorithm A
%the value $b = d_v-1$ is fixed, while in algorithm B it can vary between $\left\lceil d_v/2 \right\rceil$
%and $d_v-1$ during decoding (we denote by $\left\lceil \cdot \right\rceil$ the ceiling function). 
%Algorithm A is simpler to implement, but algorithm B ensures better performance.

We have already observed that, differently from algebraic hard-decision codes, the decoding radius of LDPC codes is generally unknown.
%So, numerical simulations are usually performed to have a precise estimate of their
%error correction capability.
So, numerical simulations are usually exploited for estimating the performance, but
such approach is
time demanding and unpractical for the purpose of dimensioning the QC-LDPC code-based
cryptosystem.
In the following, we show how to estimate the performance of the BF algorithm, when 
applied in the considered scenario, through theoretical arguments that are very similar to those 
developed in \cite{Zarrinkhat2004}.
%We must notice that, in our case, the BF algorithm works on the Tanner graph of an LDPC code,
%but it is commonly accepted that the theoretical performance evaluation based on the
%tree representation of the code still provides reliable estimations.

Let us suppose that Bob, after having received the ciphertext, performs decoding through Algorithm A.
At each iteration of the algorithm, we denote by $p^{cc}$ the probability that a bit is not
in error and a generic parity-check equation evaluates it correctly. Instead, $p^{ci}$ is the 
probability that a bit is not in error and a parity-check equation evaluates it incorrectly.
Similarly, $p^{ic}$ and $p^{ii}$ are the probabilities that a bit is in error and a 
parity-check equation evaluates it correctly and incorrectly, respectively.
In the considered context, by using simple combinatorial arguments, it is possible to verify that the following expressions hold:

\begin{equation}
\left\{ 
\begin{array}{l}
p^{cc}\left(q_l\right) = \sum_{\begin{subarray}{c} j=0\\ j \ \mathrm{even} \end{subarray}}^{\min\left\{d_c-1,q_l\right\}}\frac{{d_c-1 \choose j}{n-d_c \choose q_l-j}}{{n-1 \choose q_l}} \\
p^{ci}\left(q_l\right) = \sum_{\begin{subarray}{c} j=0\\ j \ \mathrm{odd} \end{subarray}}^{\min\left\{d_c-1,q_l\right\}}\frac{{d_c-1 \choose j}{n-d_c \choose q_l-j}}{{n-1 \choose q_l}} \\
p^{ic}\left(q_l\right) = \sum_{\begin{subarray}{c} j=0\\ j \ \mathrm{even} \end{subarray}}^{\min\left\{d_c-1,q_l\right\}}\frac{{d_c-1 \choose j}{n-d_c \choose q_l-1-j}}{{n-1 \choose q_l-1}} \\
p^{ii}\left(q_l\right) = \sum_{\begin{subarray}{c} j=0\\ j \ \mathrm{odd} \end{subarray}}^{\min\left\{d_c-1,q_l\right\}}\frac{{d_c-1 \choose j}{n-d_c \choose q_l-1-j}}{{n-1 \choose q_l-1}} \\
\end{array} 
\right.,
\label{eq:CheckProbabilities}
\end{equation}
where $d_c = n_0 \cdot d_{v}$ is the row weight of the matrix $\mathbf{H}$ and $q_l$ is the average number
of residual errors after the $l$-th iteration. It must be $q_0 \leq t = t'm$;
we fix $q_0 = t = t'm$ in order to obtain worst-case estimates (maximum error propagation).

Let us suppose that, after the $l$-th iteration, the estimate of a bit is in error. Based on \eqref{eq:CheckProbabilities}, 
we can calculate the probability that, during the subsequent iteration, the message originating from its corresponding
variable node is correct; this can be expressed as:

\begin{equation}
f^{b}\left(q_l\right)=\sum^{d_v-1}_{j=b}{d_v-1 \choose j} {\left[p^{ic}\left(q_l\right)\right]}^j{\left[p^{ii}\left(q_l\right)\right]}^{d_v-1-j}.
\label{eq:fb}
\end{equation}
Similarly, the probability of incorrectly evaluating, in a single iteration of the algorithm, a bit that is
not in error can be expressed as:
\begin{equation}
g^{b}\left(q_l\right)=\sum^{d_v-1}_{j=b}{d_v-1 \choose j} {\left[p^{ci}\left(q_l\right)\right]}^j{\left[p^{cc}\left(q_l\right)\right]}^{d_v-1-j}.
\label{eq:gb}
\end{equation}

Under the ideal assumption of a cycle-free Tanner graph
(which implies to consider an infinite-length code), 
the average number of residual bit errors at the $l$-th iteration, $q_l$, results in:
\begin{equation}
q_l = t - t \cdot f^{b}\left(q_{l-1}\right) + \left(n-t\right) \cdot g^{b}\left(q_{l-1}\right).
\end{equation}
Based on this recursive procedure, we can calculate a waterfall threshold
%of a regular cycle-free Tanner graph with variable nodes degree $d_v$ and check nodes degree $d_c$, 
by finding the 
maximum value $t = t_{\mathrm{th}}$ such that $\displaystyle \mathop{{\rm lim}}_{l\to \infty }\left(q_l\right)=0$.

Actually, different values of $t_{\mathrm{th}}$ can be found by
different choices of $b$. So, rather than resorting only to Algorithm A (in which $b = d_v-1$ is fixed),
we can also optimize the choice of $b$ by looking for the
minimum $t_{\mathrm{th}}$ for each $b \in \left\{\left\lceil d_v/2 \right\rceil, \ldots, d_v-1\right\}$.
This way, variants of Algorithm A with better choices of $b$ can be obtained. For each set of code parameters,
we will refer to the optimal choice of $b$ in the following.

Table \ref{tab:BFdecoding} reports the threshold values, so obtained, for several values
of the circulant block size $p$, code rates $2/3$ ($n_0=3$) and $3/4$ ($n_0=4$), and two
values of column weight: $d_v=13, 15$.
%We remind that the code length can be obtained as $n = n_0 \cdot p$. The choice of $b$ has been optimized as described above.

\begin{table}[ht]
\renewcommand{\arraystretch}{1.3}
\caption{Threshold values for BF decoding with fixed (optimal) $b$.}
\label{tab:BFdecoding}
\centering
\small\begin{tabular}{|@{\hspace{0.8mm}}c@{\hspace{0.8mm}}|@{\hspace{0.8mm}}c@{\hspace{0.8mm}}||c@{\hspace{0.8mm}}|@{\hspace{0.8mm}}c@{\hspace{0.8mm}}|@{\hspace{0.8mm}}c@{\hspace{0.8mm}}|@{\hspace{0.8mm}}c@{\hspace{0.8mm}}|@{\hspace{0.8mm}}c@{\hspace{0.8mm}}|@{\hspace{0.8mm}}c@{\hspace{0.8mm}}|@{\hspace{0.8mm}}c@{\hspace{0.8mm}}|@{\hspace{0.8mm}}c@{\hspace{0.8mm}}|@{\hspace{0.8mm}}c@{\hspace{0.8mm}}|@{\hspace{0.8mm}}c@{\hspace{0.8mm}}|@{\hspace{0.8mm}}c@{\hspace{0.8mm}}|@{\hspace{0.8mm}}c@{\hspace{0.8mm}}|@{\hspace{0.8mm}}c@{\hspace{0.8mm}}|}
\hline
%& & \multicolumn{13}{c|}{$p$ [bits]} \\
\multicolumn{2}{|c||}{$p$ [bits]} & $4096$ & $5120$ & $6144$ & $7168$ & $8192$ & $9216$ & $10240$ & $11264$ & $12288$ & $13312$ & $14336$ & $15360$ & $16384$ \\
\hline
\hline
\multirow{2}{*}{$n_0=3$} 	& $d_v=13$ & $190$ & $237$ & $285$ & $333$ & $380$ & $428$ & $476$ & $523$ & $571$ & $619$ & $666$ & $714$ & $762$ \\
													& $d_v=15$ & $192$ & $240$ & $288$ & $336$ & $384$ & $432$ & $479$ & $527$ & $575$ & $622$ & $670$ & $718$ & $766$ \\
\hline
\multirow{2}{*}{$n_0=4$} 	& $d_v=13$ & $181$ & $225$ & $270$ & $315$ & $360$ & $405$ & $450$ & $495$ & $540$ & $585$ & $630$ & $675$ & $720$ \\
													& $d_v=15$ & $187$ & $233$ & $280$ & $327$ & $374$ & $421$ & $468$ & $515$ & $561$ & $608$ & $655$ & $702$ & $749$ \\
\hline
\end{tabular}
\end{table}

%The limit of such theoretical analysis is that it refers to asymptotic conditions, that is, assuming infinite code lengths and cycle-free Tanner graphs.
In more realistic scenarios, 
with finite code lengths and closed loops in the Tanner graphs, and also adopting a finite number 
of decoding iterations, there is no guarantee that the error rate is arbitrarily small for 
$t \leq t_{\mathrm{th}}$. In this sense, the values in Table \ref{tab:BFdecoding} should be seen as an 
optimistic assumption.
%To counterbalance the considerations above, 
However, we can observe that the performance 
achievable by BF with fixed $b$ can be improved in a number of ways.

One of these improvements has been mentioned above, and consists in 
using Algorithm B (i.e., variable $b$). On the other hand, more recently, 
the original Gallager algorithms have been made more efficient through 
further, and more elaborated, variants \cite{Miladinovic2005, Cho2010}. %\cite{Miladinovic2005, Shan2005, Cho2010}.
%\cite{Miladinovic2005, Shan2005, Kamabe2010, Cho2010}.
Such improved versions reduce the gap in performance with respect to the SPA,
which is able to reach extremely small error
rates for values of $t$ even above the BF threshold $t_{\mathrm{th}}$ \cite{Baldi2009}.
So, taking into account these aspects, we can consider the BF threshold values as reliable 
approximations of the decoding radius of the considered QC-LDPC codes.

As concerns complexity, we can estimate the number of binary operations needed for each 
iteration of the algorithm over the code Tanner graph.
During an iteration, each check node receives $d_c$ binary values and EX-ORs them,
for a total of $d_c - 1$ binary sums. The result is then EX-ORed again with the message coming
from each variable node before sending it back to the same node, thus requiring further
$d_c$ binary sums. So, the total number of operations at check nodes is $r (2d_c - 1)$.
Similarly, each variable node receives $d_v$ check sum values and counts the number
of them that are unsatisfied; this requires $d_v$ operations.
After that, for each neighboring check node, any variable node updates the number of unsatisfied check
sums by excluding the message received from that node and compares the result with the 
threshold $b$; this requires further $2d_v$ operations.
So, the total number of operations at variable nodes is $n (3d_v)$.
In conclusion, the cost of one iteration of bit flipping can be estimated as
\begin{equation}
C_{\mathrm{BF}}^{(1)} = r \left(2d_c - 1\right) + n \left(3d_v\right) = 5nd_v - r.
\label{eq:CBFi}
\end{equation}

Based on \eqref{eq:CBFi}, and considering the computational effort required for calculating the $\mathbf{x}\cdot\mathbf{Q}$ 
and $\mathbf{u}'\cdot\mathbf{S}$ products, we can estimate
the total cost, in terms of binary operations, for each decrypted bit.
The values obtained %, for the choices of $p$, $n_0$ and $d_v$ we have considered before, 
are reported in Table \ref{tab:Decryption}, where $m = 7$ has been assumed and a BF algorithm with 10 average iterations has been considered.

\begin{table}[ht]
\renewcommand{\arraystretch}{1.3}
\caption{Binary operations needed for each decrypted bit by using BF decoding.}
\label{tab:Decryption}
\centering
\small\begin{tabular}{|@{\hspace{1mm}}c@{\hspace{1mm}}|@{\hspace{1mm}}c@{\hspace{1mm}}||c@{\hspace{1mm}}|@{\hspace{1mm}}c@{\hspace{1mm}}|@{\hspace{1mm}}c@{\hspace{1mm}}|@{\hspace{1mm}}c@{\hspace{1mm}}|@{\hspace{1mm}}c@{\hspace{1mm}}|@{\hspace{1mm}}c@{\hspace{1mm}}|@{\hspace{1mm}}c@{\hspace{1mm}}|@{\hspace{1mm}}c@{\hspace{1mm}}|@{\hspace{1mm}}c@{\hspace{1mm}}|@{\hspace{1mm}}c@{\hspace{1mm}}|@{\hspace{1mm}}c@{\hspace{1mm}}|@{\hspace{1mm}}c@{\hspace{1mm}}|@{\hspace{1mm}}c@{\hspace{1mm}}|}
\hline
\multicolumn{2}{|c||} {$p$ [bits]} & $4096$ & $5120$ & $6144$ & $7168$ & $8192$ & $9216$ & $10240$ & $11264$ & $12288$ & $13312$ & $14336$ & $15360$ & $16384$ \\
\hline
\hline
\multirow{2}{*}{$n_0=3$} 	& $d_v=13$ & $1476$ & $1544$ & $1611$ & $1668$ & $1726$ & $1784$ & $1827$ & $1899$ & $1928$ & $2014$ & $2014$ & $2130$ & $2101$ \\
													& $d_v=15$ & $1626$ & $1694$ & $1761$ & $1818$ & $1876$ & $1934$ & $1977$ & $2049$ & $2078$ & $2164$ & $2164$ & $2280$ & $2251$ \\
\hline
\multirow{2}{*}{$n_0=4$} 	& $d_v=13$ & $1598$ & $1694$ & $1790$ & $1877$ & $1963$ & $2050$ & $2107$ & $2223$ & $2252$ & $2396$ & $2381$ & $2569$ & $2511$ \\
													& $d_v=15$ & $1731$ & $1828$ & $1924$ & $2010$ & $2097$ & $2183$ & $2241$ & $2356$ & $2385$ & $2529$ & $2515$ & $2702$ & $2644$ \\
\hline
\end{tabular}
\end{table}

By using the same parameters, and considering $v=6$ quantization bits for the decoder
messages, we have estimated the decryption complexity with SPA decoding \cite{Baldi2009};
the results are reported in Table \ref{tab:DecryptionSPA}.
To decode by using the SPA guarantees the best error correction performance at 
the threshold value $t = t_{\mathrm{th}}$. However, in comparison with Table \ref{tab:Decryption}, 
the adoption of BF decoding gives a significant advantage over the SPA in terms of 
decryption complexity.

\begin{table}[ht]
\renewcommand{\arraystretch}{1.3}
\caption{Binary operations needed for each decrypted bit by using SPA decoding.}
\label{tab:DecryptionSPA}
\centering
\small\begin{tabular}{|@{\hspace{1mm}}c@{\hspace{1mm}}|@{\hspace{1mm}}c@{\hspace{1mm}}||c@{\hspace{1mm}}|@{\hspace{1mm}}c@{\hspace{1mm}}|@{\hspace{1mm}}c@{\hspace{1mm}}|@{\hspace{1mm}}c@{\hspace{1mm}}|@{\hspace{1mm}}c@{\hspace{1mm}}|@{\hspace{1mm}}c@{\hspace{1mm}}|@{\hspace{1mm}}c@{\hspace{1mm}}|@{\hspace{1mm}}c@{\hspace{1mm}}|@{\hspace{1mm}}c@{\hspace{1mm}}|@{\hspace{1mm}}c@{\hspace{1mm}}|@{\hspace{1mm}}c@{\hspace{1mm}}|@{\hspace{1mm}}c@{\hspace{1mm}}|@{\hspace{1mm}}c@{\hspace{1mm}}|}
\hline
%& & \multicolumn{13}{c|}{$p$ [bits]} \\
\multicolumn{2}{|c||} {$p$ [bits]} & $4096$ & $5120$ & $6144$ & $7168$ & $8192$ & $9216$ & $10240$ & $11264$ & $12288$ & $13312$ & $14336$ & $15360$ & $16384$ \\
\hline
\hline
\multirow{2}{*}{$n_0=3$} 	& $d_v=13$ & $9791$ & $9859$ & $9926$ & $9983$ & $10041$ & $10099$ & $10142$ & $10214$ & $10243$ & $10329$ & $10329$ & $10445$ & $10416$ \\
													& $d_v=15$ & $11261$ & $11329$ & $11396$ & $11453$ & $11511$ & $11569$ & $11612$ & $11684$ & $11713$ & $11799$ & $11799$ & $11915$ & $11886$ \\
\hline
\multirow{2}{*}{$n_0=4$} 	& $d_v=13$ & $9068$ & $9164$ & $9260$ & $9347$ & $9433$ & $9520$ & $9577$ & $9693$ & $9722$ & $9866$ & $9851$ & $10039$ & $9981$ \\
													& $d_v=15$ & $10375$ & $10471$ & $10567$ & $10653$ & $10740$ & $10826$ & $10884$ & $10999$ & $11028$ & $11172$ & $11158$ & $11345$ & $11288$ \\
\hline
\end{tabular}
\end{table}

%A comparison between the encryption and decryption complexity of the proposed cryptosystem and that of the original McEliece cryptosystem will be presented in the next Section.

\section{Security level}
\label{sec:Security}

Attacks can be divided into two classes:
\begin{itemize}
\item structural attacks, aimed at recovering the secret code;
\item decoding attacks, aimed at decrypting the transmitted ciphertext.
\end{itemize}

\subsection{Structural attacks}
\label{subsec:StructAttacks}

Structural attacks against the McEliece cryptosystem aim at recovering the
secret code from the public one; thus, they are strongly influenced by the
family of codes used.

The original proposal of using binary Goppa codes has still never suffered
a structural attack.
Recently, a new class of distinguishers has been proposed for high rate 
McEliece cryptosystems \cite{Faugere2011}.
They allow to distinguish the generator matrix of a Goppa code from a randomly
picked binary matrix, under the condition that the code rate is very high (i.e., close to $1$).
Though this is an important result, it does not concern most Goppa-based instances
of the McEliece cryptosystem, in which code rates below $0.8$ are used.

With the aim of reducing the size of the public key, it has been attempted many
times to replace binary Goppa codes with other families of codes.
In order to meet the target, the codes must have a structured nature, which
allows achieving compact matrix representations, but, on the other hand,
may be exploited by suitably devised structural attacks.

Two of the most interesting proposals of this type are those considering
QC and QD codes \cite{Berger2009, Misoczki2009}, which are able to achieve 
strong reductions in the key size with respect to classical Goppa codes.
The codes used in these systems are still algebraic codes which fall in
the class of alternant codes.
A first attack against these variants has been presented in \cite{Umana2010},
and exploits linear redundancies in subfield subcodes of GRS codes.
An even more effective attack procedure against them has been proposed
in \cite{Faugere2010, Faugere2010a}, and exploits a system of bi-homogeneous 
polynomial equations which hold for alternant codes.
The structured nature of the QC and QD codes used in \cite{Berger2009} and
\cite{Misoczki2009} results in a set of highly structured algebraic equations,
which allow to mount an efficient key-recovery attack.

The system we consider is based on codes which are basically designed at
random, apart from the need to avoid short cycles in their Tanner graphs.
So, they do not have any algebraic structure, which prevents structural
attacks of this kind.

%Other structural attacks have been devised against different families of
%codes, like the Sidelnikov and Shestakov attack against the GRS codes used
%in the Niederreiter cryptosystem \cite{Sidelnikov1992}, the Minder and Shokrollahi
%attack against Reed-Muller codes \cite{Minder2007}, and Overbeck's attack
%against rank-metric codes \cite{Overbeck2008}, but they do not apply to the
%considered system, due to the different nature of the codes used.

The proposed system is also immune against the new class of distinguishers proposed
in \cite{Faugere2011}.
In fact, the transformation there proposed cannot be applied to the QC-LDPC codes
of the type we consider, due to the lack of algebraic structure.
Although the existence of a distinguisher cannot be considered as a proof of weakness,
the non-existence is a further argument in favor of the robustness of the QC-LDPC
code-based cryptosystem.

The most dangerous structural attacks against the considered cryptosystem come
from the existence of a sparse matrix representation for the private code.
In fact, if one tries to exploit the sparse nature of LDPC codes for reducing
the public key size, density reduction attacks can be mounted, which are able to recover
the private parity-check matrix \cite{Monico2000, Baldi2007ICC, Baldi2006}.
Even though the sparse representation is hidden for the public code, but the
latter is still permutation equivalent to the private code, it may be recovered
by an attacker through a search for low weight codewords in the dual of the public
code \cite{Baldi2007ICC}.
So, when using LDPC codes, the public code must not be permutation equivalent
to the private one, as instead occurs in the original McEliece cryptosystem.
We have shown in \cite{Baldi2009, Baldi2008} that this can be achieved by
using the transformation matrix $\mathbf{Q}$ in the place of a permutation
$\mathbf{P}$ for computing the public key.

As already shown in Section \ref{sec:System}, the matrix $\mathbf{Q}$ must be sparse in
order to allow correcting all intentional errors.
If also the matrix $\mathbf{S}$ is chosen to be sparse \cite{Baldi2007ISIT},
a structural attack still exists \cite{Otmani2008}, which is able to recover
a sparse representation for the secret code.
However, it suffices to choose a dense matrix $\mathbf{S}$, as in the original
McEliece cryptosystem, to avoid such attack \cite{Baldi2008}.

\subsection{Decoding attacks}
\label{subsec:DecAttacks}

Due to the low weight ($t'$) of the intentional error vector, decoding attacks 
against the considered system are often more dangerous than structural attacks,
and provide the smallest WF.
These attacks aim at solving the decoding problem, that is, obtaining the error 
vector $\mathbf{e}$ used for encrypting a ciphertext. 
A way for finding $\mathbf{e}$ is to search for the minimum weight codewords of an 
extended code, generated by:
\begin{equation}
\mathbf{G}''=\left[\begin{array}{c}\mathbf{G}'\\\mathbf{x}\end{array}\right].
\end{equation}

The WF of such attacks can be determined by referring to the Stern's algorithm \cite{Stern1989}. More precisely, we have used an updated version of this algorithm \cite{Peters2010}, that results in minimum WF for the class of codes here considered. It must be said that several advances have recently appeared in the literature for improving the running time of the best decoding algorithms for binary random codes (see \cite{May2011}, \cite{Becker2012}, for example). These papers, however, often aim at evaluating the performance of information set decoding in asymptotic conditions, i.e., for codes with infinite length, while we prefer to rely on actual operation counts, which are not reflected in these recent works.
Another recent advance in this direction is represented by ``ball collision decoding'' \cite{Bernstein2011}, which is able to achieve important WF reductions asymptotically.
However, for finite code lengths and security levels even above those of interest here, such improvement is negligible \cite{Bernstein2011}.

On the other hand, we must observe that, in the QC-LDPC code-based cryptosystem, a further speedup is obtained by considering
that, because of the quasi-cyclic property of the codes, each block-wise cyclically shifted 
version of the ciphertext $\mathbf{x}$ is still a valid ciphertext.
So, the eavesdropper can continue extending $\mathbf{G}''$ by adding block-wise
shifted versions of $\mathbf{x}$, and can search for one among as many shifted 
versions of the error vector.
So, in order to estimate the minimum WF,
we have 
considered the optimum number of shifted ciphertexts that can be used by an attacker in the
generator matrix of the extended code.

For each QC-LDPC code, we have calculated the maximum number of intentional errors 
$t' = \left\lfloor t/m \right\rfloor$ by considering $m=7$ and the estimated error
correction capability $t$ reported in Table \ref{tab:BFdecoding}.
The values obtained are reported in Table \ref{tab:PublicCodeErrors}.

The minimum WF values, obtained in such conditions, are shown 
in Table \ref{tab:SecLevel}.
For $n_0=3$, the WF of the attack to the dual code, also based on the improved version of Stern's
algorithm, is about $2^{161}$, when $d_v=13$, and $2^{184}$, when $d_v=15$. So, we have
reported the former of such values in Table \ref{tab:SecLevel} for those cases in which the decoding 
attack WF would be higher.
The same has been done for $n_0=4$, for which the WF of the attack to the dual code
is about $2^{154}$ and $2^{176}$ for $d_v=13$ and $d_v=15$, respectively.

\begin{table}[ht]
\renewcommand{\arraystretch}{1.3}
\caption{Number of intentional errors $t'$ introduced by Alice.}
\label{tab:PublicCodeErrors}
\centering
\small\begin{tabular}{|@{\hspace{0.8mm}}c@{\hspace{0.8mm}}|@{\hspace{0.8mm}}c@{\hspace{0.8mm}}||c@{\hspace{0.8mm}}|@{\hspace{0.8mm}}c@{\hspace{0.8mm}}|@{\hspace{0.8mm}}c@{\hspace{0.8mm}}|@{\hspace{0.8mm}}c@{\hspace{0.8mm}}|@{\hspace{0.8mm}}c@{\hspace{0.8mm}}|@{\hspace{0.8mm}}c@{\hspace{0.8mm}}|@{\hspace{0.8mm}}c@{\hspace{0.8mm}}|@{\hspace{0.8mm}}c@{\hspace{0.8mm}}|@{\hspace{0.8mm}}c@{\hspace{0.8mm}}|@{\hspace{0.8mm}}c@{\hspace{0.8mm}}|@{\hspace{0.8mm}}c@{\hspace{0.8mm}}|@{\hspace{0.8mm}}c@{\hspace{0.8mm}}|@{\hspace{0.8mm}}c@{\hspace{0.8mm}}|}
\hline
\multicolumn{2}{|c||}{$p$ [bits]} & $4096$ & $5120$ & $6144$ & $7168$ & $8192$ & $9216$ & $10240$ & $11264$ & $12288$ & $13312$ & $14336$ & $15360$ & $16384$ \\
\hline
\hline
\multirow{2}{*}{$n_0=3$} 	& $d_v=13$ & $27$ & $33$ & $40$ & $47$ & $54$ & $61$ & $68$ & $74$ & $81$ & $88$ & $95$ & $102$ & $108$ \\
													& $d_v=15$ & $27$ & $34$ & $41$ & $48$ & $54$ & $61$ & $68$ & $75$ & $82$ & $88$ & $95$ & $102$ & $109$ \\
\hline
\multirow{2}{*}{$n_0=4$} 	& $d_v=13$ & $25$ & $32$ & $38$ & $45$ & $51$ & $57$ & $64$ & $70$ & $77$ & $83$ & $90$ & $96$ & $102$ \\
													& $d_v=15$ & $26$ & $33$ & $40$ & $46$ & $53$ & $60$ & $66$ & $73$ & $80$ & $86$ & $93$ & $100$ & $107$ \\
\hline
\end{tabular}
\end{table}

\begin{table}[ht]
\renewcommand{\arraystretch}{1.3}
\caption{Security level of the QC-LDPC code-based cryptosystem for $m=7$.}
\label{tab:SecLevel}
\centering
\small\begin{tabular}{|@{\hspace{0.8mm}}c@{\hspace{0.8mm}}|@{\hspace{0.8mm}}c@{\hspace{0.8mm}}||c@{\hspace{0.8mm}}|@{\hspace{0.8mm}}c@{\hspace{0.8mm}}|@{\hspace{0.8mm}}c@{\hspace{0.8mm}}|@{\hspace{0.8mm}}c@{\hspace{0.8mm}}|@{\hspace{0.8mm}}c@{\hspace{0.8mm}}|@{\hspace{0.8mm}}c@{\hspace{0.8mm}}|@{\hspace{0.8mm}}c@{\hspace{0.8mm}}|@{\hspace{0.8mm}}c@{\hspace{0.8mm}}|@{\hspace{0.8mm}}c@{\hspace{0.8mm}}|@{\hspace{0.8mm}}c@{\hspace{0.8mm}}|@{\hspace{0.8mm}}c@{\hspace{0.8mm}}|@{\hspace{0.8mm}}c@{\hspace{0.8mm}}|@{\hspace{0.8mm}}c@{\hspace{0.8mm}}|}
\hline
%& & \multicolumn{13}{c|}{$p$ [bits]} \\
\multicolumn{2}{|c||} {$p$ [bits]} & $4096$ & $5120$ & $6144$ & $7168$ & $8192$ & $9216$ & $10240$ & $11264$ & $12288$ & $13312$ & $14336$ & $15360$ & $16384$ \\
\hline
\hline
\multirow{2}{*}{$n_0=3$} 	& $d_v=13$ & $2^{54}$ & $2^{63}$ & $2^{73}$ & $2^{84}$ & $2^{94}$ & $2^{105}$ & $2^{116}$ & $2^{125}$ & $2^{135}$ & $2^{146}$ & $2^{157}$ & $2^{161}$ & $2^{161}$ \\
													& $d_v=15$ & $2^{54}$ & $2^{64}$ & $2^{75}$ & $2^{85}$ & $2^{94}$ & $2^{105}$ & $2^{116}$ & $2^{126}$ & $2^{137}$ & $2^{146}$ & $2^{157}$ & $2^{168}$ & $2^{179}$ \\
\hline
\multirow{2}{*}{$n_0=4$} 	& $d_v=13$ & $2^{60}$ & $2^{73}$ & $2^{85}$ & $2^{98}$ & $2^{109}$ & $2^{121}$ & $2^{134}$ & $2^{146}$ & $2^{153}$ & $2^{154}$ & $2^{154}$ & $2^{154}$ & $2^{154}$ \\
													& $d_v=15$ & $2^{62}$ & $2^{75}$ & $2^{88}$ & $2^{100}$ & $2^{113}$ & $2^{127}$ & $2^{138}$ & $2^{152}$ & $2^{165}$ & $2^{176}$ & $2^{176}$ & $2^{176}$ & $2^{176}$ \\
\hline
\end{tabular}
\end{table}

In order to give an example of system design, we can consider the parameters of the Goppa code suggested 
in \cite{Bernstein2008} for achieving $80$-bit security (i.e., $\mathrm{WF} = 2^{80}$), that are: $n=1632$, $k=1269$
and $t = 33$.
Under a suitable CCA2-secure conversion, they give a key size of $57581$ bytes for both the McEliece cryptosystem
and the Niederreiter version.
The encryption and decryption complexities, estimated through the formulas in \cite[p. 27]{Canteaut1996},
result in $817$ and $2472$ operations per bit, respectively, for the McEliece 
cryptosystem and $48$ and $7890$ operations per bit for the Niederreiter version.
The transmission rate is $0.78$ for the McEliece cryptosystem and $0.63$ for the
Niederreiter version.

A similar security level can be reached by the QC-LDPC code-based cryptosystem
with $n_0 = 4$, $p = 6144$ and $d_v = 13$.
%, even considering the possible speedup 
%deriving from the adoption of improved versions of Stern's algorithm.
In this case, as reported in Table \ref{tab:KeySize}, the public key size is $2304$ bytes, i.e.,
$25$ times smaller than in the Goppa code-based McEliece and Niederreiter cryptosystems.
The transmission rate is $0.75$, similar to that of the Goppa code-based McEliece cryptosystem
and higher than in the Niederreiter version.
The encryption and decryption complexities, as reported in Tables \ref{tab:Encryption}
and \ref{tab:Decryption}, result in $1206$ and $1790$
operations per bit, respectively. So, the complexity increases in the encryption stage, 
but, by exploiting the BF algorithm, the decryption complexity 
is reduced.
%with respect to the McEliece cryptosystem and the Niederreiter version

So, we can conclude that, for achieving the same security level, the QC-LDPC
code-based cryptosystem can adopt smaller keys and comparable or
higher transmission rates with respect to the classical Goppa code-based McEliece and Niederreiter 
cryptosystems.
Moreover, this does not come at the expense of a significantly increased complexity.

\section{Conclusion}
\label{sec:Conclusion}

We have deepened the analysis of a variant of the McEliece cryptosystem using
QC-LDPC codes in the place of Goppa codes.
Such modification is aimed at overcoming the main drawbacks of the original system,
while still allowing to reach a satisfactory security level.

We have proposed to adopt bit flipping algorithms for decoding the QC-LDPC codes,
in such a way as to achieve a rather good performance while strongly reducing the decoding 
complexity with respect to the SPA.
The adoption of bit flipping decoding has also allowed to develop simple analytical
tools for estimating the error correction capability of the considered codes, thus
simplifying the system design by avoiding the need for long numerical simulations.
Together with the methods we have described to evaluate complexity, these tools
provide the system designer with a fast procedure for optimizing the choice of the
cryptosystem parameters.

\section*{Acknowledgment}
The authors wish to thank Rafael Misoczki for having suggested improvements in Table \ref{tab:BFdecoding}, and Ludovic Perret for fruitful discussion on the complexity of the attacks to the cryptosystems based on QD codes and QC-LDPC codes.

They are also grateful to the editor and the anonymous reviewers, who provided valuable comments and suggestions
which significantly helped to improve the quality of this paper.

\newcommand{\BIBdecl}{\setlength{\itemsep}{0.01\baselineskip}}

\end{document}